\documentclass[12pt]{iopart}

\usepackage[latin1]{inputenc}
\usepackage{graphicx}

\newcommand{\beq}{\begin{equation}}
\newcommand{\eeq}{\end{equation}}
\newcommand{\dd}{{\rm d}}
\newcommand{\ld}{\lambda_D}
\newcommand{\rr}{{\mathbf r}}
\newcommand{\ek}{{\mathcal E}_k}

\begin{document}
\article{tutorial}{About the dynamics and thermodynamics of trapped ions}

\author{C. Champenois}
 \address{ Physique
des Interactions Ioniques et Mol\'eculaires, UMR 6633, CNRS and 
Universit\'e de Provence, Centre de Saint J\'er\^ome, Case C21,
13397 Marseille Cedex 20, France}

\ead{caroline.champenois@univ-provence.fr}

\begin{abstract}
This tutorial introduces the dynamics of charged particles in a radiofrequency trap in a very general manner to point out the differences between the dynamics in a quadrupole and in a multipole trap. When  dense samples are trapped, the dynamics is modified by the Coulomb repulsion   between ions. To take into account this repulsion, we propose to use a method, originally developed for particles in Penning trap, that model the ion cloud as a cold fluid. This method can not reproduce the organisation of cold clouds as crystals but it allows one to scale the size of  large samples with the trapping parameters and the number of trapped ions, for different linear geometries of trap.
\end{abstract}

\pacs{52.27.Jt, 52.25.Kn, 37.10.Pq}

\section{Introduction}
Quadrupole radiofrequency traps appear to be the obvious choice for most applications where charged particles need to be trapped, cooled and manipulated by laser interactions. Optical frequency metrology and quantum information processing give examples of such a control of the ion motion by atom-laser interaction as shown in several contributions to this special issue. Up to now, all the protocols make use of the harmonic nature of the potential induced by the radiofrequency field and were demonstrated with one or few ions. On the other hand, higher order  radiofrequency traps are efficiently used with large samples where buffer gas cooling is preferred to laser cooling. This is the case, for example, for microwave frequency standards working with  large samples of Hg$^+$ ions \cite{prestage01,prestage07} or with experiments studying cold reactions with molecules \cite{teloy74,mikosch04}. In the prospect of new experiments where one could think of applying quantum information processing methods to a larger sample (of atomic or molecular ions), it may be useful to consider using higher order radiofrequency traps than the usual quadrupole trap. The first purpose of this tutorial is to give a general background about the dynamics of ions in radiofrequency traps to point out the differences between the dynamics inside a quadrupole and a multipole trap (here we call multipole trap a linear radiofrequency trap with a higher symmetry order than a quadrupole). The second objective of this tutorial is to present a mean-field approach already used for particles in Penning traps and called the {\it nonneutral plasma}. Using this method, it is possible to scale the size of a trapped sample with the relevant physical parameters of the problem.  This general approach is useful to understand what is so special about the quadrupolar geometry even when large samples are trapped. This tutorial is organised in two parts, the first one describes the dynamics of single charged particle in a radiofrequency field of very general geometry. The second part introduces the mean field approach that allows one to take into account the Coulomb repulsion in large samples. This model is used in the linear geometry to establish relations between the trapping parameters and the characteristics of the trapped sample.

\section{General description of radiofrequency trapping \label{sec_dyn}}
\subsection{The most general case\label{sec_dyngene}}
Let's  try to get some insights about the dynamics of trapped ions (mass $m$ and charge $q$) starting with a non specific geometry. The development presented here is greatly inspired from a broad review written by D. Gerlich about multipoles and their use to study ion-molecule reaction dynamics \cite{gerlich92}. Let's assume that the potential between a set of electrodes can be written as the sum of a  static voltage  $\Phi_s(\mathbf{r})$ and a radiofrequency voltage $\Phi_{rf}(\mathbf{r},t)$ oscillating at frequency $\Omega$ and responsible for the effective trapping of the charged particle. The motion of a single ion is then governed by the local electric field
\beq
\mathbf{E}(\mathbf{r},t)=\mathbf{\nabla}\Phi_s(\mathbf{r})+\mathbf{\nabla}\Phi_{rf}(\mathbf{r},t)= \mathbf{E_s}(\mathbf{r})+\mathbf{E_0}(\mathbf{r})\cos(\Omega t).
\eeq
To compute the trajectory of this ion, the difficulty lies in the integration of the equations of motion in a time varying electric field. So we first assume that there is no static electric field ($\mathbf{E_s}(\mathbf{r})=0$). If the electric field were homogeneous, the motion would  revert to the rf-driven oscillation
\beq
\mathbf{r}(t)=-\frac{q\mathbf{E_0}}{m\Omega^2}\cos(\Omega t)=-\mathbf{a}\cos(\Omega t),
\eeq
where $\mathbf{a}$ is the amplitude of this oscillation. As  $\mathbf{E_0}(\mathbf{r})$ is non homogeneous, this amplitude varies in space and the motion can be split into two contributions: the motion driven by the rf field $\mathbf{R_1}(t)=-\mathbf{a}(t)\cos(\Omega t)$ and a slower motion  $\mathbf{R_0}(t)$ induced by the  variation of the amplitude of the radiofrequency field. The local electric field can then be expanded as
\beq 
 \mathbf{E_0}(\mathbf{r}(t))= \mathbf{E_0}(\mathbf{R_0})-(\mathbf{a}(t).\mathbf{\nabla})\mathbf{E_0}(\mathbf{R_0})\cos(\Omega t)+\cdots \label{eq_E}
\eeq
The resolution proposed here for  the dynamical equations relies on two  approximations
\begin{enumerate}
\item the first order expansion of $ \mathbf{E_0}$ around $\mathbf{R_0}$ in Eq.~({\ref{eq_E}}).
\item {\it the adiabatic approximation} which assumes that the typical evolution time scale of $\mathbf{a}$ and $\mathbf{\dot{R}_0}$ is far longer than the rf period. 
\end{enumerate}
In this frame of approximations, the slow dynamics is determined by
\beq
 m\mathbf{\ddot{R}_0}=-\frac{q^2}{4 m \Omega^2} \mathbf{grad}(E_0^2)
\eeq
and, as far as $\mathbf{R}_0(t)$ is concerned, it is as if the particle is trapped in a static potential well, called the {\it pseudo}-potential:
\beq\label{eq_pseusoV}
V^*(\mathbf{r})=\frac{q^2 E_0^2(\mathbf{r})}{4 m \Omega^2}.  
\eeq
Building a pseudopotential well implies $\mathbf{grad}(E_0^2(\mathbf{r}))>0$ in the three directions. In practice, the fulfilment of this condition  is not sufficient to assure the stability of the ion's trajectory.  A conservative stability criterion is given by a local adiabaticity criterion, introduced by Teloy and Gerlich \cite{teloy74,gerlich92}, and defined by the relative variation of $\mathbf{E_0}$ seen over the amplitude of the driven motion oscillation:
\beq
\eta_{ad}=\frac{|2(\mathbf{a}.\mathbf{\nabla})\mathbf{E_0}|}{|\mathbf{E_0}|}=\frac{2 q |\mathbf{\nabla}\mathbf{E_0}|}{m\Omega^2}. 
\label{eq_eta}
\eeq
This adiabaticity parameter depends on the location in the trap except for the particular case of quadratic radiofrequency voltage for which $\mathbf{\nabla}\mathbf{E_0}$ is homogeneous. In the multipole case, numerical simulations  and experimental observations mentioned by D. Gerlich \cite{gerlich92}  show that the empirical limit $\eta_{ad}< 0.3$ guarantees adiabaticity in most cases. A more recent experimental study of the loss mechanism in a 22-pole trap \cite{mikosch08} complementary to a more general model of effective trapping volume for multipole \cite{mikosch07} has demonstrated stability up to $\eta_{ad}<0.36 \pm 0.02$. A safe criterion for stability could then be $\eta_{ad}(R_{max}) < 0.34 $.

If the static electric field is switched on again, according to the superposition theorem, the motion is governed by the effective static potential
\beq\label{effV}
V^*(\mathbf{r})=\frac{q^2 E_0^2(\mathbf{r})}{4 m \Omega^2}+q\Phi_s(\mathbf{r}) . 
\eeq
The rf-driven motion (also called micromotion) $\mathbf{R_1}(t)$ is deduced from the motion in the pseudopotential (called macromotion):
\beq
\mathbf{R_1}(t)=-\frac{q \mathbf{E_0}(\mathbf{R_0})}{m \Omega^2} \cos(\Omega t)
\label{eq_micro}
 \eeq
and its amplitude  is proportionnal to the local rf-electric field. As this electric field must increase with $ \mathbf{R_0}$ for effective trapping, the micromotion amplitude increases with the distance from the trap center. As this driven motion  can not be cooled, the only way to reduce this amplitude is to keep the ion at the node of the electric field.

\subsection{Focus on the linear trap}
In most experiments requiring a precise control of the dynamics and kinetic energy of the ions, micromotion is a side effect that can bring a large contribution to the Doppler effect and reduces the precision reached on the kinetic energy of the ions. In a linear geometry where a radiofrequency voltage is applied only in the transverse plane  with a translation symmetry along the remaining axis, the node of the radiofrequency electric field is the symmetry axis. When large samples are trapped, it  gives rise to a reduced micromotion compared to spherical traps, as the ions are closer to the node of the electric field. In this section, we assume such a linear geometry where the trap consist of $2k$ equally spaced rods as electrodes for transverse trapping and where the confinement along the axis of symmetry is reached by a static potential (the geometry of the electrodes used for the axial confinement is not very relevant in this tutorial).

\begin{figure}
\begin{center}$
\begin{array}{ccc}
\includegraphics[width=5.5cm]{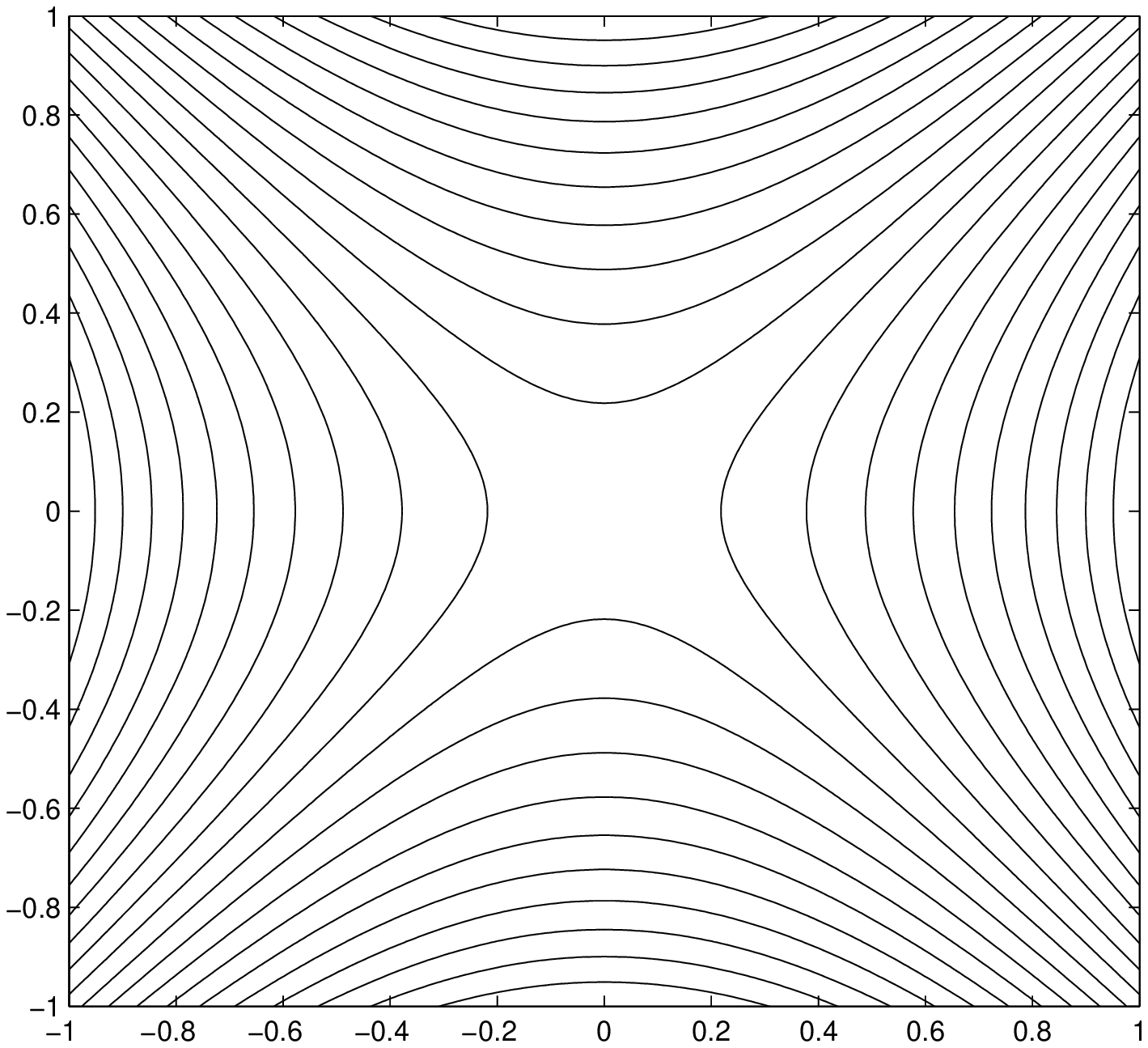} &
\includegraphics[width=5.5cm]{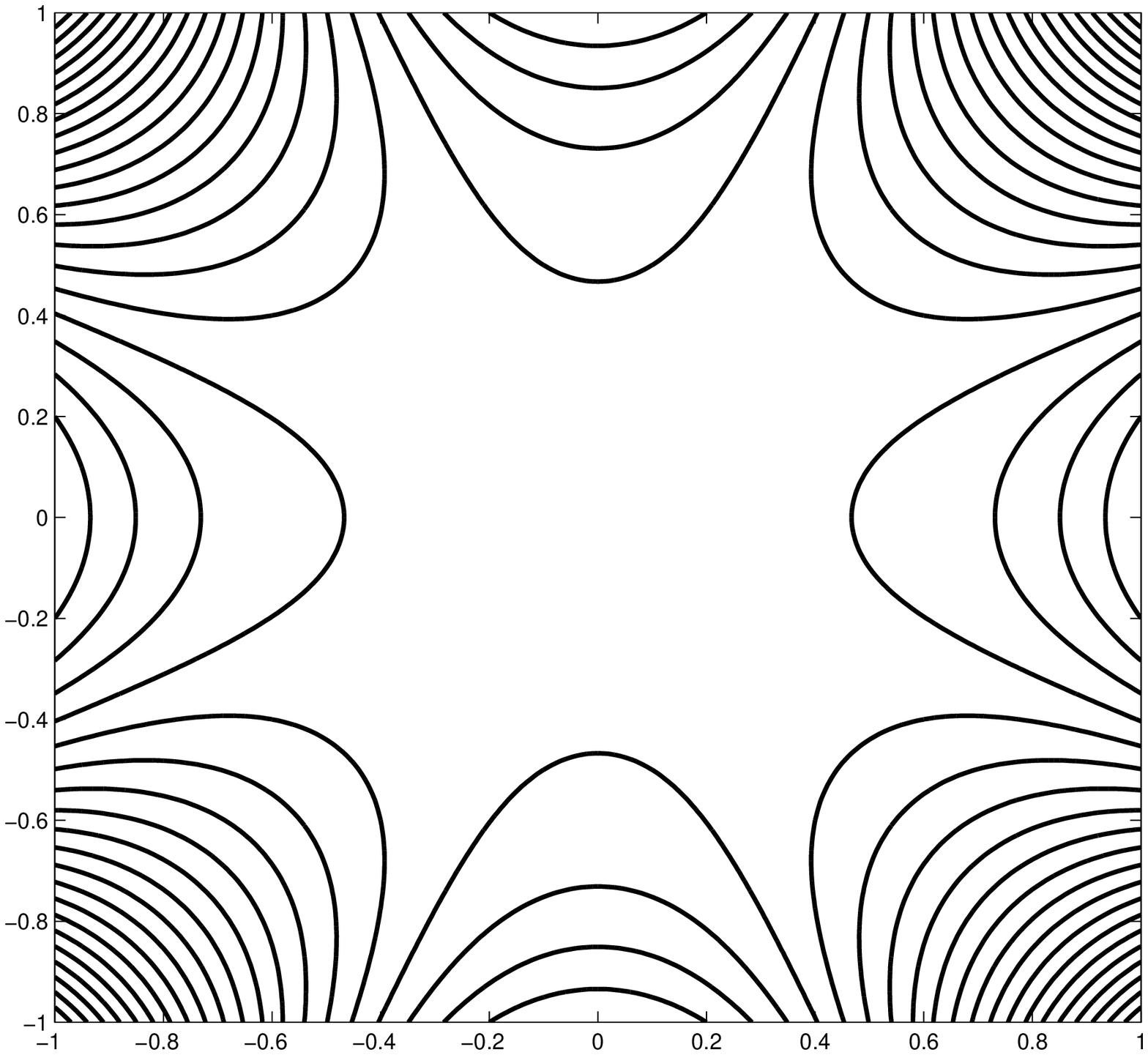}&
\includegraphics[width=5.5cm]{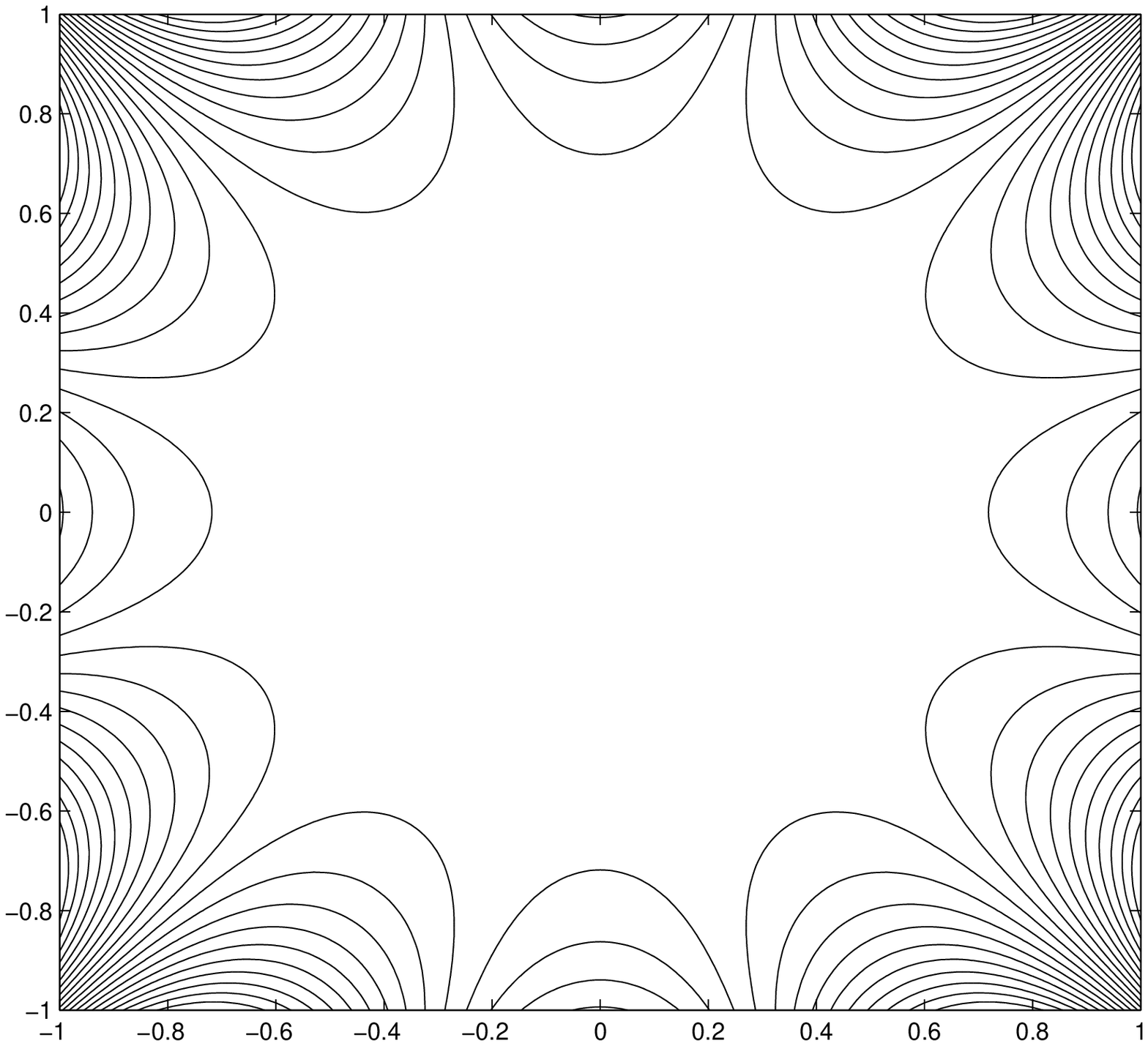}
\end{array}$
\end{center}
\caption{equipotential curves in the plane of a $2k$-pole with a spatial dependance of the rf potential defined by $ (r/r_0)^k \cos (k\theta) $. From left to right, $2k=4, 8, 12$. \label{fig_2kpoleV}}
\end{figure}

As a first step, we ignore this static potential and assume that  voltages $+V(t)/2=U_s/2-V_0/2 \cos(\Omega t)$ and $-V(t)/2$ are applied on alternate electrodes  (so the voltage difference between neighbouring electrodes is $V(t)$, notice that in Gerlich's paper this voltage is chosen to be $2 V(t)$). The $2k$ electrodes are rods located at the same distance $r_0$ from the trap axis and solutions of the Laplace equation $\Delta \Phi=0$ are linear combinations of 
\beq
\Phi_k(\mathbf{r},t)=\Phi_0(t) (r/r_0)^k \cos (k\theta) 
\eeq
where $(r,\theta)$ are the polar coordinates in the $(x,y)$ plane. The contribution of each order $k$ is fixed by the boundary conditions defined by the shape and position of the electrodes. Figure \ref{fig_2kpoleV} shows the equipotential lines in the transverse plane of an ideal multipole where only one contribution of defined symmetry order defines the potential. This potential distribution can be realised ideally if the electrode surfaces exactly match the equipotential surfaces. 
Deriving the radiofrequency electric field generated by such a potential, the equations governing the motion in the plane of a $2k$-pole are
\begin{eqnarray}
\ddot{x}/r_0 - F_k(t)(r/r_0)^{k-1} \cos(k-1) \theta  &= & 0  \label{eq_dynMP} \\ 
\ddot{y}/r_0 + F_k(t)(r/r_0)^{k-1} \sin(k-1) \theta  &= & 0  \nonumber
\end{eqnarray}
with $F_k(t)=kqU_s/(2mr_0^2)-k qV_0/(2mr_0^2)\cos (\Omega t)$.  As an example, let's have a look at the equations of motion of an ion in an octupole ($2k=8$):
\begin{eqnarray}
\ddot{x} + F_4(t) (x^3-3y^2 x)/r_0^2  &= & 0 \nonumber\\ 
\ddot{y} - F_4(t) (y^3-3x^2 y)/r_0^2  &= & 0. 
\end{eqnarray}
These equations are non-linear and coupled. The non-linearity makes the stability of individual trajectories sensitive to their initial conditions. As a consequence, it is not possible to define  absolute stability conditions based only on the working parameters of the trap. An estimation  of the local stability is provided by the adiabaticity criterion (Eq.~(\ref{eq_eta})). For a $2k$-pole, it increases like $r^{k-2}$ and is uniform only in the quadrupolar case ($2k=4$):
\beq
\eta_{ad}=k(k-1)\frac{qV_0r^{k-2}}{m\Omega^2r_0^k}.
\label{eq_etaMP}
\eeq

 In a quadrupole, and only in a quadrupole, the equations of motion take the simple forms: 
\begin{eqnarray}
\ddot{x} - F_2(t) x  &= & 0 \nonumber\\ 
\ddot{y} + F_2(t) y  &= & 0  
\end{eqnarray}
which are linear and uncoupled for $x$ and $y$.  The introduction of the reduced time scale $\xi=\Omega t/2$ leads to the well-known Mathieu equation
\beq
\frac{\dd^2 u}{\dd \xi^2}+(a_u-2q_u \cos(\Omega t))u(\xi)=0
\label{mathieu}
\eeq
$(u=x,y)$, which belongs to the  family of differential equations with periodic coefficients. Solutions of this equation can be found in many  textbooks \cite{book_ghosh, book_werth}. The  solutions have the form 
\beq
u(t)=A e^{i \omega_u t} \sum_n C_{2n}e^{in\Omega t}+ B e^{-i \omega_u t} \sum_n C_{2n}e^{-in\Omega t}
\eeq
with $\omega_u=\beta_u \Omega/2$, $\beta_u$ depending only on the Mathieu parameters $a_u$ and $q_u$ through continuous fractions. $a_u$ and $q_u$ depend on the trapping parameters by:
\begin{eqnarray}
q_x=\frac{2qV_0}{m \Omega^2 r_0^2}  & ; &   q_y=-\frac{2qV_0}{m \Omega^2 r_0^2} \\ 
a_x=-\frac{4qU_s}{m \Omega^2 r_0^2} &  ; &  a_y=\frac{4qU_s}{m \Omega^2 r_0^2}.
\end{eqnarray}
These solutions are stable if $0 \leq \beta_u \leq 1$. This condition defines  areas in the plane defined by ($a_u$,$q_u$): the stability regions. In practice and because of technical limitations, radiofrequency traps are operated in the lowest stability region (the one including (0,0)).

The solutions of the Mathieu equation can be expanded in the lowest order approximation which requires $(|a_u|, q_u^2) \ll 1$ and implies that $\beta_u \ll 1$. In this limit $ \beta_u=\sqrt{a_u+q_u^2/2}$ and
\beq
u(t)=U\cos(\omega_u t)\left(1+\frac{q_u}{2}\cos \Omega t \right).
\eeq
This is exactly what we would obtain using the pseudopotential well resulting from the adiabatic approximation. It says that the main motion (or macromotion) is a harmonic oscillation of frequency $\omega_x$, perturbed by the driven radiofrequency motion (or micromotion). The amplitude of this micromotion is proportional to the harmonic oscillation amplitude and to the Mathieu equation parameter $q_u$.

The Mathieu equations, or more generally the equations of motion given by Eq (\ref{eq_dynMP}), are strictly true for a single ion in an ideal trap and do not include the Coulomb repulsion between ions. When several ions are trapped, the correlations between ions (often called space charge effect in its mean field approach) can be ignored for warm enough or very dilute samples where the averaged Coulomb repulsion is low compared to the kinetic energy. The coupling parameter $\Gamma=q^2/(4\pi \epsilon_0 a k_BT )$, the ratio of the average nearest-neighbor Coulomb repulsion energy and the thermal energy, quantifies this competition between correlation and thermal motion (the usual definition sets $a$  as the Wigner-Seitz radius and is related to the density by $4\pi n a^3/3=1$). For low $\Gamma$ ($\Gamma \ll 1$), the Coulomb repulsion can be ignored and it is relevant to use the single ion equations of motion to learn about the ions dynamics. With laser cooling techniques, $\Gamma$ as high as few hundred can be reached. As soon as $\Gamma \geq 1$, correlations can not be ignored to calculate dynamics or equilibrium properties of the trapped ions. This problem is widely addressed by two methods. The first one is based  on computer simulations of the experiment by calculating the position and velocity of each ion by a Monte-Carlo method \cite{pollock73}, when only the statistical equilibrium is studied, or by molecular-dynamics simulations \cite{rahman86,dubin88,prestage91} when information about the dynamics is also wanted. These computer simulations are an efficient tool to calculate the crystal structure appearing in trapped ions as soon as $\Gamma \geq 100$, depending on the trapping parameter, the number of ions and the temperature \cite{dubin88, schiffer00, drewsen98, zhang07}. It is even possible to include buffer gas or laser Doppler cooling in the model, by an effective damping force \cite{prestage91, schiller03} or by a stochastic force induced by the momentum kicks in each collision or absorption/emission process \cite{walther93}. In this tutorial I wish to present another method which has been developed for electrons or ions in Penning traps but which is very relevant for ions in radiofrequency trap, using a model for the atomic sample called the  {\it nonneutral plasma}.  This model  uses a mean-field approach where the ion-ion Coulomb repulsion is taken into account by the total field created by the charge distribution. One can show \cite{dubin99} that the equilibrium state of such a system in a harmonic pseudopotential is equivalent to that of a One Component Plasma (OCP), a model system where the charged particles are embedded in a neutralizing background charge. It allows one to calculate the density and size of trapped charged samples in the dense and cold limit, a requirement which fits very well with laser cooled atoms. Contrary to the molecular dynamics or Monte Carlo methods, it can not reproduce  the internal Coulomb crystal structure observed in experiments but it is of great help to scale the size of a linear trap with the desired number of ions. This is the question that I try to answer in the following section. Most of the concepts introduced here can be found in the review by D.H.E Dubin and T.M. O'Neil \cite{dubin99} concerning the thermal equilibrium states of trapped nonneutral plasmas. 

\section{The mean-field approach\label{sec_mf}}

When a large sample is trapped, a mean field approach can be used to study the global behaviour of the sample. This treatment  has been mostly used in the context of Penning traps, for electrons or ions \cite{book_davidson,ichimaru82} but also in 3D radiofrequency  quadrupole traps (or Paul traps) \cite{cutler86}. In this model, a thermal equilibrium state is assumed and the sample is treated like a cold fluid. This approximation holds if the plasma size is large compared to the Debye length $\ld$. This length characterises the  scale over which a  mean field approach is relevant to treat the Coulomb repulsion. It depends on the density $n$ and temperature $T$ of the plasma as $\ld=\sqrt{k_B T \epsilon_0/q^2 n}$ and decreases for cold and dense plasmas. The cold fluid model allows one also to calculate the aspect ratio (radius over length) of a cloud trapped in a harmonic potential \cite{turner87}. These predictions have been confirmed experimentally with laser cooled ions in a Penning trap \cite{brewer88} and in a linear rf quadrupole in the isotropic  \cite {hornekaer01} and anisotropic \cite{frohlich05} case.

 In the context of radiofrequency trap, the equilibrium state concept requires us to use the pseudopotential approach to represent the trapping potential. Like in the previous section, we assume a linear geometry but the same points developed here could be applied to 3D multipole trap.  To find an experimental characterisation of the thermal equilibrium of an atomic cloud  trapped in a 3D octopole trap, the reader should refer to \cite{walz94} where the density profile is studied by means of laser induced fluorescence resolved in space. In the linear geometry, the pseudopotential in the transverse plane of an ideal multipole is defined by (see Eq (\ref{effV}))
\beq
V^*(\mathbf{r})=\frac{q^2V_0^2}{32 \ek}\left(\frac{r}{r_0}\right)^{2k-2}+\frac{qU_s}{2}\left(\frac{r}{r_0}\right)^{k} \cos k \theta.
\eeq
where $\ek=m\Omega^2 r_0^2/(2 k^2)$ is a characteristic energy. In the following, we consider that there is no static voltage $U_s$ applied to the rods but we take into account  the static potential required for axial confinement and its effect on  the transverse pseudopotential. The total effective potential becomes
\beq 
V^*(\mathbf{r})=\frac{q^2V_0^2}{32 \ek}\left(\frac{r}{r_0}\right)^{2k-2} +\frac{q\kappa V_{end}}{2z_0^2}(2z^2-r^2),
\eeq
where $\kappa$ is a loss factor depending on the geometry of the end electrodes relative to the rods. It includes all screening effects that can explain the reduction between the potential applied on the end electrodes $V_{end}$ and the one effectively seen by the ions.
For $2k=4$, the shape of the transverse pseudopotential remains quadratic and no major impact is expected from the transverse deconfining effect of the axial confinement. On the contrary, for $2k=8, 12 \ldots$, the shape of the pseudopotential is modified. The axis of the trap becomes an unstable position and the potential minimum is shifted to $r=r_{min}$ defined by 
\beq
r_{min}^{2k-4}=\frac{r_0^{2k-2}}{z_0^2}\frac{16 \ek \kappa V_{end}}{(k-1)q V_0^2}.
\eeq
We see later that this shift of the potential minimum from the center of the trap has no effect on the density distribution calculated in  the mean-field model.

According to the work of  Dubin, Driscoll, O'Neil and Prasad \cite{oneil79, prasad79, dubin99} and assuming the ergodic hypothesis, the thermal equilibrium state of a nonneutral plasma with non negligible correlations  can be described by the $N$-particle Gibbs distribution for plasmas as small as $N\geq 100$ (for smaller $N$, fluctuations are too big to identify the average over the microcanonical and the canonical-or Gibbs-distribution). Integrated over the $N$ velocities and  $N-1$ positions, the Gibbs distribution gives access to the plasma density distribution $n(\rr)$:
\begin{equation}
n(\rr) = {\mathcal N} \exp\left[-\frac{{\mathcal E}(\rr)}{k_B T}\right]
\end{equation}
where ${\mathcal N}$ is a normalisation constant, and ${\mathcal E}(\rr)$ is the energy of a particle at position~$\rr$. It results from the contribution of the confining potential $ V^*(\mathbf{r})=q\phi_T(\rr)$ and the mean-field Coulomb repulsion potential $q\phi_q(\rr)$ created by all the charges surrounding point $\rr$ (the image charge effect is neglected as we assume the charges are far enough from any conducting surface \cite{dubin93}). The normalisation issue can be solved by defining the density with respect to the density in the center of the trap $n({\mathbf 0})=n_0$ :
\begin{eqnarray}
n(\rr) & = & n_0 \exp [\Psi(\rr)] \\
\Psi(\rr) &  = & -\frac{q}{k_BT}\left(\phi_T(\rr) + \phi_q(\rr) -\phi_T({\mathbf 0})- \phi_q({\mathbf 0})\right).
\label{eq_defPsi}
\end{eqnarray}
The density distribution and the potential created by the charges are related by the Poisson equation
\beq
\Delta \phi_q(\rr)=-\frac{qn(\rr)}{\epsilon_0}.
\label{eq_poisson}
\eeq
This system of equations is then self-consistent but one can get rid of some normalisation difficulties by studying only the logarithmic density profile $\Psi(\rr)$ \cite{oneil79, prasad79} and using Eq.~(\ref{eq_defPsi},\ref{eq_poisson}), one can show that
\beq
\Delta\Psi(\rr)=  \frac{q^2 n_0}{k_B T \epsilon_0}\left[\exp[\Psi(\rr)]-\frac{\epsilon_0}{q n_0}\Delta\phi_T\right].
 \label{eq_dif}		
\eeq
It is obvious from Eq.~ (\ref{eq_dif}) that the Debye length relative to the central density $\lambda_D=\sqrt{k_B T \epsilon_0/(q^2 n_0)}$ will be the relevant length scale for the density profile. This equation also shows that the static contribution to the effective potential $V^*$ has no impact on the density profile as this contribution obeys the Laplace equation $\Delta \phi=0$ and that only the pseudopotential associated with the radiofrequency field controls this profile. Eq.~(\ref{eq_dif}) also points out the difference between quadrupoles and multipoles, regarding the density profile of a nonneutral plasma. Indeed, in the  cold fluid limit ($T\rightarrow 0$), to prevent divergence of the density, $\exp[\Psi(\rr)]-\frac{\epsilon_0}{q n_0}\Delta\phi_T$ must also tend to 0 which means that $n(\rr) \rightarrow \epsilon_0 \Delta\phi_T/q$. This equation can also be deduced from the condition for mechanical equilibrium  of a shell of ions for which  the trapping field  balances the field created by the charges inside the shell.  In our radiofrequency multipole context, this becomes
\beq
\lim_{T\to 0} n(r)=\frac{ \epsilon_0 (k-1)^2 V_0^2}{8 \ek r_0^2}\left(\frac{r}{r_0}\right)^{2k-4}
\label{eq_densite}
\eeq
 which is uniform only for the quadrupolar geometry. For higher order geometry, the density is expected to increase with the distance from the center of the trap, leading to an empty center and a geometry like a tube. This is confirmed by molecular dynamics simulations presented in \cite{okada07} for calcium ions in an octopole.

 We now want to use the differential equation (\ref{eq_dif})	to scale a cloud size. It appears that the situation is very different for a quadrupole from other multipole geometries. So we first deal with quadrupole geometry, which is very close to the Penning trap configuration, and then expand our method to higher order geometry.

\subsection{In a linear quadrupole\label{sec_lq}}
In the particular case of a linear quadrupole, the pseudopotential can be written as
\beq
\phi_T(r,z)=\frac{m}{2q}\omega_r^2 r^2 + \frac{m}{2q}\omega_z^2 z^2
\eeq
with $\omega_r^2=\omega_x^2-\omega_z^2/2$, $\omega_x$ being defined by the solution of the Mathieu equation \cite{book_ghosh}. As mentioned above, the contribution determining the density profile is
\beq
\Delta \phi_T(r,z)=2\frac{m}{q}\omega_x^2 
\eeq
and does not depend on the static end voltage. The second term in the differential equation (\ref{eq_dif}) is then a constant $2m\epsilon_0\omega_x^2/(q^2 n_0)$ that can be understood  as the ratio of the uniform limit density  $n_c=2m\epsilon_0\omega_x^2/q^2$ that is reached for low temperature, divided by the realised central density $n_0$. This ratio is bigger than 1 as the  limit density  $n_c$ is the highest reachable density for a given harmonic potential. In the following this ratio $n_c/n_0$ is expressed as $\gamma+1$, $\gamma$ measuring how far the system is from the uniform limit density ($\gamma >0$). Finally, the differential equation determining the density distribution is
\beq
\Delta \Psi(r,z)=\frac{1}{\lambda_D^2}\left[ \exp[\Psi(r,z)] - \gamma-1 \right]
\eeq
To simplify the integration, we introduce the reduced coordinates $(\rho, \xi)=(r/\lambda_D,z/\lambda_D)$ and solve:
\beq
\Delta \Psi(\rho,\xi)= \exp[\Psi(\rho,\xi)] - \gamma-1.
\eeq

{\bf Case of a prolate cloud $L \gg R$}: In the following, we assume  the cloud is prolate and  the dependence of the density with $z$ is negligible compared to its dependence with $r$. This simplifies the problem to the integration of a one dimension differential equation
\beq
\frac{1}{\rho}\frac{\partial \Psi}{\partial \rho} +\frac{\partial^2 \Psi}{\partial \rho^2}=\exp [\Psi(\rho)]-\gamma-1.
\label{eq_difrho}
\eeq
To have an insight into the role of $\gamma$ in this equation, it is useful to look for an approximate solution, valid close to the trap center. For $\Psi(\rho) \ll 1$, $\exp [\Psi(\rho)]-1 \simeq \Psi(\rho)$. In this approximation, the solution of the differential equation (\ref {eq_difrho}) is $\gamma (1-I_0(\rho))$ where $I_0(\rho)$ is the modified Bessel function of order 0. Using the lowest order expansion of $I_0(\rho)$, one can deduce that for $\rho^2 \ll 4/\gamma$, $n(\rho) \simeq n_0(1-\gamma \rho^2/4)$. So the smaller  $\gamma$ is, the larger (in reduced parameter) is the range where the density profile is nearly flat.

Integration of Eq (\ref{eq_difrho}) (with the limit conditions $\Psi(0)=0$ and $\Psi^{'}(0)=0$) results in the profile $\Psi(\rho)$ and its exponential $n(\rho)/n_0$ which implicitly depends on $\gamma$. To get real values for the number of trapped ions, the absolute density profile $n(r)$ and the size of the sample, one has to find other relations between the relevant functions and parameters:

\begin{itemize}
\item First, in the frame of our approximation, the total number of ions is 
\beq
N=\int \int  n(r) 2\pi r \dd r \dd z=2L  \int  n(r) 2\pi r \dd r
\eeq
if $L$ is the half length of the cloud. By using the reduced parameter $\rho$ and the definition of $\lambda_D$, one can show that
\beq
\frac{N}{2L}=\frac{k_BT \epsilon_0}{q^2}\int \exp [\Psi(\rho)] 2\pi \rho \dd \rho.
\label{eq_N}
\eeq
This number of ions depends implicitly on the central density $n_0$ through the parameter $\gamma=n_c/n_0-1$. Here, the number of ions per unit length $N/2L$, the temperature $T$  and the central density $n_0$ are the three relevant parameters for the density profile problem and only two of them are free as they are related by eq (\ref{eq_N}).
\item Second, to relate the two physical free parameters  $N/2L$ and $T$ to the real size of the cloud, it is useful to notice that, even if $\lambda_D$ does not appear explicitly in the differential equation, it is not a free parameter since
\beq
\frac{\lambda_D^2}{\gamma+1}=\frac{k_B T}{2m\omega_x^2 }.
\label{eq_couple}
\eeq
\end{itemize}

\begin{figure}[ht]
\centerline{\includegraphics[width=12cm]{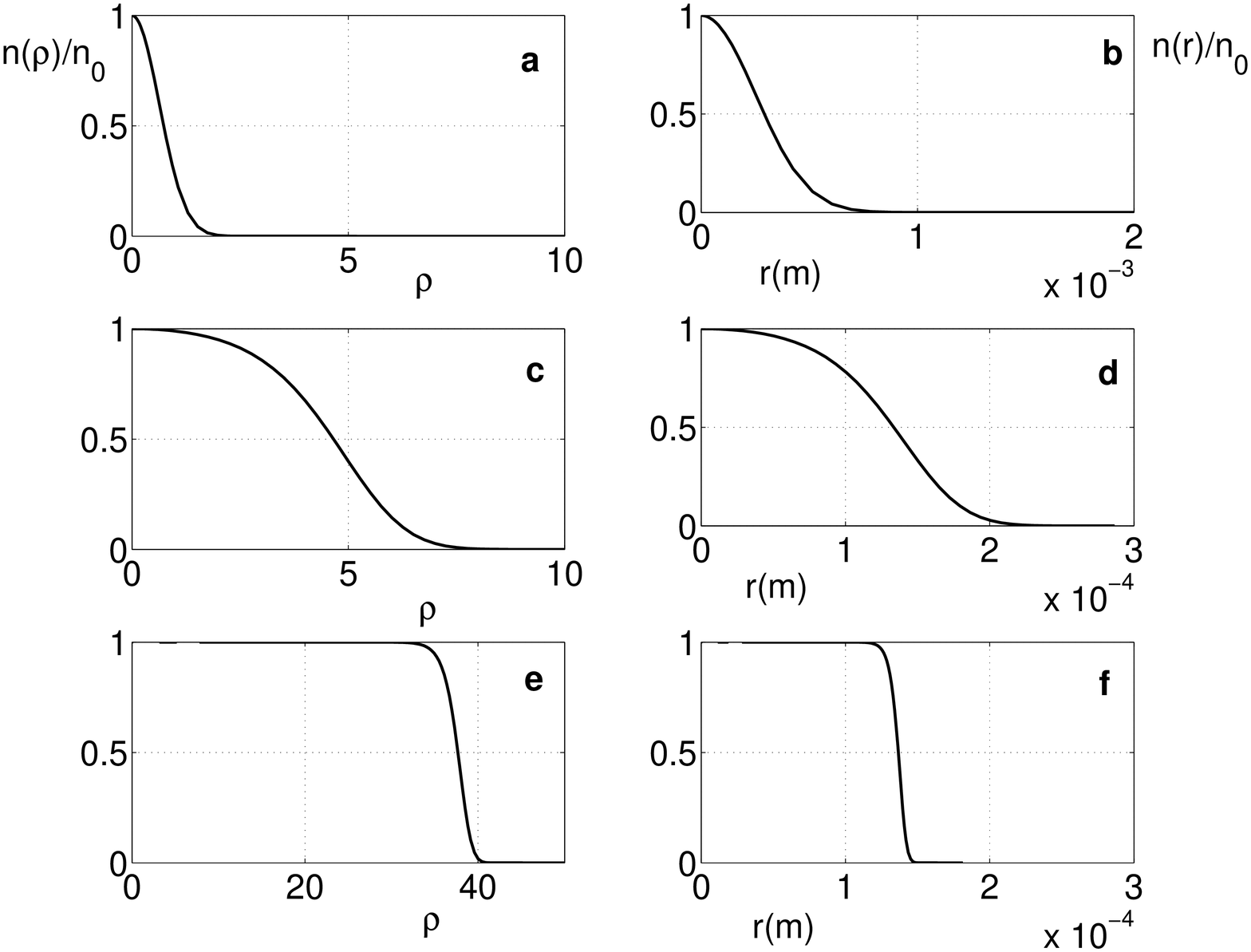}}
 \caption{ Density profiles versus the reduced radius $\rho$ ({\bf a,c,e}) or the radius $r$ ({\bf b,d,f}) for a prolate cloud with $10^5$ ions per mm, at different temperatures ({\bf a,b}: T=10 000 K, $\gamma=5$, {\bf c,d}: T=300 K, $\gamma=0.04$, {\bf e,f}: T=5 K, $\gamma=10^{-15}$). The scaling factor $\lambda_D$ between $\rho$ and $r$  is calculated for Ca$^{+}$ ions in a trap with $\omega_x/2\pi=1$ MHz. Notice that the x-axis of figure {\bf b} has a different scale from {\bf d} and {\bf  f}, and that the x-axis of figure {\bf e} has a different scale from {\bf a} and {\bf  c} .  }
 \label{fig_profils}
\end{figure}

As an experimentalist, one  may want to know what the size of an ion cloud will be. The solution is there: for a given temperature  $T$ and a given number of ions per unit length $(N/2L)$, $\gamma$ has to be found to fulfil the condition fixed by equation (\ref{eq_N}). Then $\lambda_D$ is deduced from equation (\ref{eq_couple}) and the radial size of the cloud is given by $R=\lambda_D \rho_{max}$. The strength of the harmonic pseudopotential $\sqrt{m}\omega_x$  plays a role only in this scaling. It becomes necessary to know its value only when the size of the cloud is required.  Examples of integrated radial profiles in the reduced and full scale coordinate are shown on figure \ref{fig_profils} for the same number of ions per unit length and same potential well $m\omega_x^2$ but for different temperatures. This figure illustrates the differences between warm and dilute cloud where $\rho_{max} \simeq 1$ ($T=10 000$ K, case {\bf a} and  {\bf b}) and a  colder and denser cloud   where $\rho_{max}\gg 1$ ($T=5$ K, case {\bf e} and  {\bf f}),where the cold fluid model is relevant.

As mentioned above, theoretically, it is  not appropriate to model warm and dilute samples by the cold fluid model, nevertheless, we can use the radial size calculated in this condition as an indication and compare it to the radius of the same sample cooled down. This is very important  for experimental issue, as the actual number of laser cooled trapped ions can be limited by the size it takes just after or during  the ionisation process.  Indeed, when ions are created by electron bombardement, their initial temperature can reach an order of magnitude of 10 000 K and comparison of plots (b,d,f) of figure \ref{fig_profils} shows that the radial size of the same sample  at 10 000K is nearly 4 times the size at 300 K and 6 times the one at 5 K. This illustrates one of the advantage of photoionisation over electron bombardement to ionise a neutral beam. During the photoionisation \cite{kjaergaard00}, the ions can be continuously loaded and laser cooled and the sample never reach as high temperature as by electron bombardement. Then the cloud remains smaller and more ions  should be loaded in a trap, with less radiofrequency heating \cite{hornekaer02, ryjkov05}.

For a  laser cooled nonneutral plasma ($N/2L \ge 10^3$/mm, $T \le 1$ K), or for a room temperature  dense sample  ($N/2L \ge 10^5$/mm, $T= 300$ K),  $\gamma$ is negligible in  Eq (\ref{eq_couple}) which  can be simplified to
\beq
\lambda_D^2=\frac{k_B T}{2m\omega_x^2}.
\eeq
This shows that in the cold fluid limit, for a given set $(T, N/2L)$, the size of a cylindrical cloud scales as $(m\omega_x^2)^{-1/2}$.

For temperatures close to the Doppler limit, the density profile of large enough sample ($\geq 10^3$ ions/mm) is flat and expands to values of $\rho \gg 1$. We can then approximate the density profile by $\exp[\Psi(\rho)]=1$ for $\rho \le \rho_{max}$ and $\exp[\Psi(\rho)]=0$ for $\rho > \rho_{max}$. The relation between the number of ions and the temperature is then simplified to:
\beq
\frac{N}{2L}=\frac{k_BT \epsilon_0}{q^2}\pi \rho_{max}^2.
\eeq
As a consequence, the radial size $R$ of the cloud is  given by:
\beq
R=\sqrt{\frac{N}{2L}}\frac{1}{\sqrt{m}\omega_x}\sqrt{\frac{q^2}{2\pi\epsilon_0}}
\eeq
which is independent of the temperature and can be considered as the minimum limit radius $R_m$ for a sample. This limit for the size of a prolate cold cloud is equivalent to a maximum limit for the density:
\beq
\frac{N}{\pi R^2 2L} \rightarrow \frac{2m\omega_x^2 \epsilon_0}{q^2}
\eeq
which is of course the limit density $n_c$ we introduced at the beginning of the text.

To have an idea of this size, for a laser cooled nonneutral plasma of Ca$^+$ in a linear quadrupole trap, the limit radius is given by the following relation where $\omega_x/2\pi$ is in MHz:
\beq
R=\sqrt{\frac{N}{2L}}\frac{1}{\omega_x/2\pi}\times 1.31 \times 10^{-8}.
\eeq

 Another issue concerns the phase transition to crystals. In finite size samples like we are dealing with, the transition from the liquid to the crystal state is not abrupt. Molecular dynamics simulations \cite{dubin88} and experiments \cite{hornekaer02} have shown that it takes place for values of the  coupling parameter $\Gamma$ between 150 and 200. As the ions arrange themselves into concentric spheroidal shells, the diffusion inside a shell persists even for $\Gamma$ as big as 300 \cite{dubin88}. To study experimentally such phase transitions in a finite system, it is then important to make sure that $\Gamma$ as high as a few hundred can be reached. For a  laser cooled dense sample, $\Gamma$  reaches a maximum limit $\Gamma_c$ depending on the limit density $n_c$ ( $a\rightarrow (4/3\pi n_c)^{1/3}$) and the temperature reached by the cooling:
\beq
 \Gamma_c=\left(\frac{q^2}{4\pi\epsilon_0}\right)^{2/3}\frac{\left(2m \omega_x^2 \right)^{1/3}}{k_B T}.
\eeq
This relation shows that the maximum $\Gamma$ that can be reached in an experiment depends on the temperature of the sample, and the harmonic pseudopotential $m \omega_x^2$. As an example,  for calcium ions and for $\omega_x/2\pi$ in MHz,
\beq
\Gamma_c=4.8 \frac{ (\omega_x/2\pi)^{2/3}}{T}.
\eeq  
Several experiments \cite{molhave00, roth05} with laser cooled ions in linear radiofrequency traps have shown that because of heating processes like  radiofrequency heating, the Doppler limit is not reached with large samples and that an optimistic estimation for the limit temperature is 10 mK. In this condition, one has to make sure to produce a high enough $\omega_x$ to be able to explore a wide range of $\Gamma$. 

All the behaviours we have just described for quadrupole linear trap can not be extrapolated to multipole traps because the cold fluid limit does not result in a uniform density but rather in a  profile density increasing with the radius (see Eq \ref{eq_densite}). Nevertheless, we can reasonably assume a thermal equilibrium and a Gibbs distribution for multipole also. Indeed, the  experimental results of the evaporation of ions from a linear 22-pole \cite{mikosch07}  are consistent with a Boltzmann statistics to characterise the velocity distribution.

\subsection{In a linear multipole\label{sec_lm}}

In this section, we use the method developed for the quadrupole linear trap in the more general case of a multipole linear trap. The differential equation governing the logarithmic density can be also written for a multipole potential, the major difference is that now the Laplacian of the pseudopotential depends on $r$:
\beq
\phi_{trap}(r,z)=\frac{ q V_0^2}{32 \ek}\left(\frac{r}{r_0}\right)^{2k-2}-\frac{m}{4q}\omega_z^2 r^2 + \frac{m}{2q}\omega_z^2 z^2.
\eeq
 With the reduced parameters $(\rho, \xi)$ as defined above, the general differential equation (\ref{eq_dif})  for a multipole is
\begin{eqnarray}
 \Delta\Psi(\rho,\xi) &=& \exp[\Psi(\rho,\xi)]-\frac{\epsilon_0}{ n_0}\frac{  V_0^2(2k-2)^2\lambda_D^{2k-4}}{32 \ek r_0^{2k-2}}\rho^{2k-4} \\
 & =& \exp[\Psi(\rho,\xi)]-\alpha \rho^{2k-4}.
\end{eqnarray}
In practice, the numerical integration depends only on $\alpha=(\epsilon_0 V_0^2(2k-2)^2\lambda_D^{2k-4})/( n_032 \ek r_0^{2k-2})$. The use of reduced parameter must not hide the limit of this method which refer the density profile to the central density $n_0$. This is certainly a limitation here as, from molecular dynamics simulations \cite{okada07}, an empty core is expected in the cold limit. We see further in the text how to interpret results obtained from this model. As the integration of $\Psi$ does not give access directly to the central density $n_0$, to scale the size of the sample one has to use the same trick as in section \ref{sec_lq}, which is to find a combination of $\alpha$ and $\lambda_D$  independent of $n_0$. From the definition of $\lambda_D$, it is easy to see that $\alpha$ scales like $n_0^{1-k}$, so the good combination is 
\beq
\frac{\lambda_D^2}{\alpha^{1/(k-1)}}=\left(\frac{32 k_BT \ek r_0^{2k-2}}{(2k-2)^2q^2 V_0^2 }\right)^{1/(k-1)}.
\label{eq_lambdaDMP}
\eeq

\begin{figure}[ht]
 \centerline{\includegraphics[width=10cm]{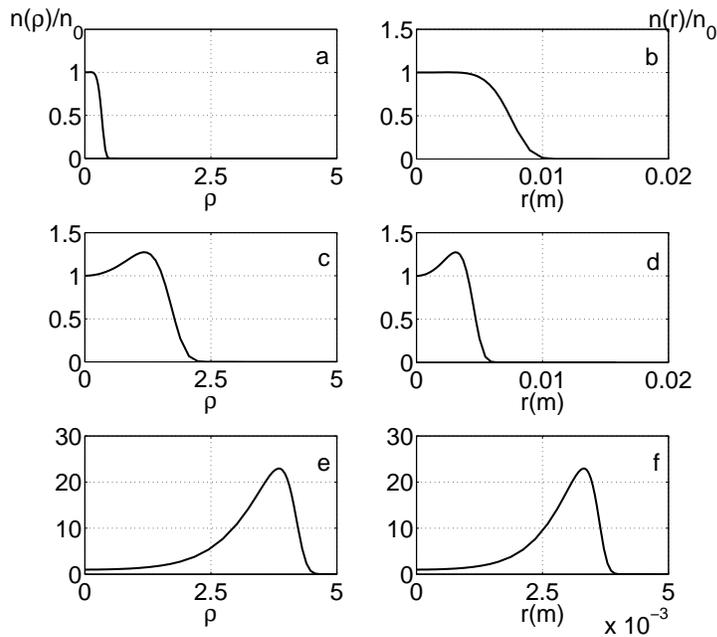}}
 \caption{  Density profiles versus the reduced radius $\rho$ ({\bf a,c,e}) or the radius $r$ ({\bf b,d,f}) for a prolate cloud with $1.6\times 10^4$ ions per mm, at different temperatures ({\bf a,b}: T=10 000 K, $\alpha=19000$, {\bf c,d}: T=300 K, $\alpha=1.8$, {\bf e,f}: T=5 K, $\alpha=0.13$). The scaling factor $\lambda_D$ between $\rho$ and $r$  is calculated for Ca$^{+}$ ions in an octopole with $r_0=1$ cm, $V_0=800$ V and $\Omega/2\pi=10$ MHz. Notice that the x-axis of figure {\bf f} has a different scale from {\bf b} and {\bf  d}.}
 \label{fig_profil8T}
\end{figure}

\begin{figure}[ht]
 \centerline{\includegraphics[width=8cm]{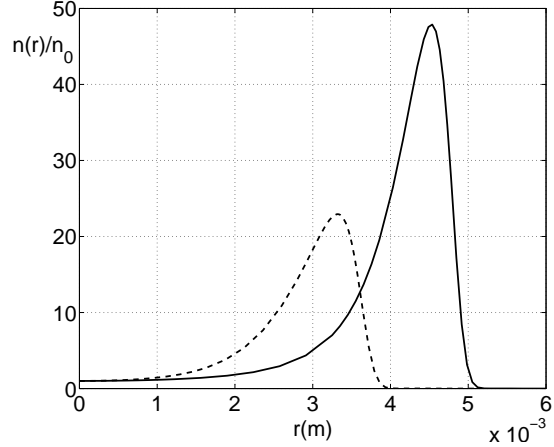}}
 \caption{Density profile of a calcium cloud of $1.6\times 10^4$ ions/mm at 5 K in two  traps of different order but with same parameters as for figure \ref{fig_profil8T} : $r_0=1$ cm, $V_0=800$ V and $\Omega/2\pi=10$ MHz. Dashed line: $k=4$, $\alpha=0.13$, $\lambda_D=0.86$ mm, $n_0=3.2 \times 10^{10}$~m$^{-3}$. Solid line:  $k=6$, $\alpha=0.0056$, $\lambda_D=1.4$ mm, $n_0=1.2 \times 10^{10}$~m$^{-3}$. }
 \label{fig_profilMP}
\end{figure}

The idea of the numerical integration is the same as in section \ref{sec_lq}. First we assume a translation symmetry along $Oz$ and do the integration along $\rho$ only. Second,  as the total number of ions per unit length is always defined by equation (\ref{eq_N}), the three coupled parameters are now $T$, $\alpha$ and $N/2L$. The smaller  $\alpha$ is, the higher is the number of ions for a given temperature. Then, the real size of the cloud is defined by $\lambda_D \rho_{max}$. Figure \ref{fig_profil8T} shows the density profile of the same cloud in the same octupole trap but for different temperature. For a matter of numerical integration, only profiles for clouds as big as $1.6\times 10^4$ ions/mm at 5 K could be calculated. For denser or colder sample, the profile is so steep that the adaptative step procedure I use for integration reached the lowest step allowed by the subroutine \cite{ode15s}. This limitation explaines why high reduced parameter $\rho$ can not be reached by the simulation, contrary to the quadrupole case. The main difference from the quadrupole geometry is visible  for the case of lower temperature  (e,f) where the density reaches a maximum more than 20 times higher than the central density. This illustrates also the limitation  we mentionned above for the cold fluid model, in comparison with molecular dynamics simulations \cite{okada07}. Choosing the density in the center of the trap $n_0$ as a reference for the density profile prevents one to  find an empty core but the high ratio between the maximum density and $n_0$ can be interpreted as a demonstration of this phenomena.  
This effect increases with the order of the multipole as can be seen on figure \ref{fig_profilMP} which shows a comparison between the density profile in a 8-pole and a 12-pole, for the same number of ions per length and the  same trapping parameters. In the higher order geometry, the radial size of the cloud is 20\% bigger than in the lower order geometry and the ratio of the maximum density over the central density is two times bigger. For the validity of our cold fluid model, it is worth noting that, by analogy with the quadrupole case, we defined the Debye length $\lambda_D$ relatively to the central density, which results into high values for $\lambda_D$, compared to the radial size of the cloud (see figure \ref{fig_profil8T} for values). Nevertheless, the relevance of our model depends on the local Debye length $\lambda_D(\rho)$ compared to the size of the sample. We can see on figure \ref{fig_profilMP} that, close to the edge of the cloud, this length $\lambda_D(\rho)$  is a lot smaller than $\lambda_D$, as the local density is far higher than $n_0$.  Consequently, using the cold fluid model to scale such a cloud seems valid in the cold and dense limit.

As for a quadrupole, the cold fluid limit allows one to estimate the minimum size of a cloud by assuming the $T \rightarrow 0$ limit. To prevent divergence of the density at this limit, $\exp[\Psi(\rho)]-\alpha\rho^{2k-4}\rightarrow 0$, with our previous notation. The maximum reduced radius visited by the ions is then determined by the linear density $N/2L$:
\beq
\frac{N}{2L}=\frac{k_BT \epsilon_0}{q^2}\frac{\alpha \pi}{k-1} \rho_{max}^{2k-2}.
\eeq
The dependance of the cold fluid limit radius $R_m$ on the trapping parameters can then be expressed as 
\beq
R_m=r_0\left(\frac{N}{2L}\frac{8\ek}{\pi \epsilon_0(k-1)V_0^2 }\right)^{1/2(k-1)}
 \label{eq_rcold}.
\eeq
To put values into this equation, let's choose the case of calcium ions in an octopole ($k=4$) and express $\Omega/2\pi$ in MHz, then
\beq
R_m=\left(\frac{N}{2L}\right)^{1/6}\left(\frac{(\Omega/2\pi) r_0^4}{V_0}\right)^{1/3}\times 0.45.
\eeq
As an example, for $V_0=400$ V, $\Omega/2\pi=1$ MHz and $r_0=1$ cm, the radial size of a cylindrical sample of $N/2L=4.2 \times 10^4$ ions/mm  is $R_m=2.4$ mm in the cold fluid limit.  As a comparison, at 300K, the numerical integration gives $R=3.8$ mm.

Also like in quadrupole traps, the set of trapping parameters is constrained by stability criterion for the trajectories of the ions. But contrary to quadrupole traps, there is no absolute criterion but one can use the local adiabaticity criterion of Teloy and  Gerlich  introduced in \ref{sec_dyngene}. For a perfect multipole like we are assuming, 
\beq
\eta_{ad}=\frac{(k-1)qV_0}{2k \ek } \left(\frac{r}{r_0}\right)^{k-2}.
\eeq
To be sure to remain in the adiabatic regime, we want to keep $\eta_{ad}\leq \eta_{lim}$ and then  the radial size $R \leq r_{max}^{ad}$ with
\beq
r_{max}^{ad} =r_0 \left(\eta_{lim}\frac{2k \ek}{(k-1)qV_0}\right)^{1/(k-2)}.
\eeq
To have an idea of how this criterion is respected, it can be useful to compare, for a given set of trapping parameters, the smallest  achievable  radial size $R_m$ and the highest allowed radius for adiabatic trajectories $r_{max}^{ad}$:
\beq
\frac{R_m}{r_{max}^{ad}}=\left(\frac{N }{L \pi \epsilon_0}\right)^{1/2(k-1)}\left(\frac{q}{k\eta_{lim}}\right)^{1/(k-2)}\left(\frac{k-1}{\ek}\right)^{k/2(k-1)(k-2)}\left(\frac{V_0}{2}\right)^{1/(k-1)(k-2)}.
\label{eq_beta}
\eeq
For Calcium ions in an octopole, this becomes
\beq
\frac{R_m}{r_{max}^{ad}}=\left(\frac{N}{2L}\frac{V_0}{\eta_{lim}^3 r_0^4\Omega^4}\right)^{1/6}\times 12.9.
\eeq
As an example, for the same parameters as above ($V_0=400$ V, $\Omega/2\pi=1$ MHz, $r_0=1$ cm) and  for the same number of ions per unit length $N/2L=4.2\times10^4$ ions/mm,  $R_m/r_{max}^{ad}=0.75$ and the adiabaticity criterion is obeyed over all the cold sample (we used $\eta_{lim}=0.3$ here).

 By looking at equations (\ref{eq_lambdaDMP},\ref{eq_rcold},\ref{eq_beta}), one can notice that if the  ratio $\Omega/V_0$ is conserved, the absolute radial size is also conserved  but not the ratio $R_m/r_{max}^{ad}$. In practice, by substituting $\Omega/V_0$ by $2\Omega/2V_0$, $R_m/r_{max}^{ad}$ is reduced by a factor $\sqrt{2}$. One can use this difference in behaviour to scale ones trap and make sure that the desired sample fits inside the adiabatic volume. 

\section{Conclusion}
In this tutorial we have described the dynamics of charged particles in a radiofrequency trap in a very general manner to point out the differences between the dynamics in a quadrupole and in a multipole trap. When  dense samples are trapped, the dynamics is modified by the Coulomb repulsion   between ions. To take into account this repulsion into the equilibrium state of the cloud, we use a method that models the ion cloud as a cold fluid. In the case of prolate clouds, it allows one to scale the size of  the samples with the trapping parameters and the number of trapped ions, for different linear geometries of trap. We think this can be useful to build an experiment where a large number of ions needs to be trapped. 

 \section*{Acknowledgement} The author would like to thank several collaborators for  their help and precious advise during the writing of this tutorial,  Masatoshi Kajita, Martina Knoop, Jofre Pedregosa, Richard Thompson and Fernande Vedel.

\newpage


\begin{thebibliography}{10}
\bibitem{ode15s}
for the figures of this paper, {I} used matlab and the {PDE} routine allowing
  the steepest profile was ode15s.
\end{thebibliography}

\end{document}